\documentstyle[12pt]{article}

\def\ttoo{\mathop{\longrightarrow}}

\begin{document}

\begin{titlepage}
\begin{flushright}
{\bf SPbU-TP-96-20}
\end{flushright}
\vskip 2cm
\begin{center}
{\LARGE\bf Tree-level $(\pi, K)$-amplitude and analyticity}
\vskip 2cm
{\Large Vladimir V. Vereshagin}
\vskip 1cm
{\it Theoretical Physics Department\\
St.Petersburg State University\\
Ulyanovskaya st 1\\
198904 St.Petersburg\\
Russia}\\[12pt]
{\it e-mail}:\\
{\tt vereshagin@phim.niif.spb.su}\\[0pt]
{\tt vergn@kom.usr.pu.ru}
\end{center}
\vfill

\begin{abstract}
\normalsize

We consider the tree-level amplitude, describing all 3
channels of the binary
($\pi ,K$)-reaction, as a
\underline{meromorphic}
polynomially bounded function of 3 dependent
\underline{complex}
variables. Relying systematically on the Mittag-Leffler
theorem, we construct 3
\underline{convergent}
partial fraction expansions, each one being applied in the
corresponding domain. Noting, that the mutual intersections
of those domains are nonempty, we realize the
\underline{analytical continuation}.
It is shown that the necessary conditions to make such a
continuation feasible, are the following: 1) The only
parameters completely determining the amplitude are the
\underline{on-shell}
couplings and masses; 2) These parameters are restricted by
a certain (infinite) system of
\underline{bootstrap}
equations; 3) The full cross-symmetric amplitude takes the
typically \underline{dual} form, the \underline{Pomeron}
contribution being taken into account; 4) This latter
contribution corresponds to a nonresonant
\underline{background},
which, in turn, is expressed in terms of cross-channel
resonances parameters.

It is demonstrated also, that the Chiral Symmetry provides
a unique scale for the mentioned parameters, the resonance
saturation effect appearing as a direct consequence of the
above results.

\end{abstract}

\end{titlepage}

\section{Introduction.}
\mbox{}

In the papers \cite{1,2}
it was demonstrated that
the requirement of ``realistic'' asymptotic
behavior (first suggested by Weinberg
\cite{3}), when applied to
$\pi\pi\rightarrow\pi\pi$ and
$\gamma\pi\rightarrow\pi\pi$
reactions amplitudes, written in the form dictated by the
large-$N_c$ limit
(\cite{4,5})
gives rise to certain (infinite) sets of self-consistency
conditions for the parameters (masses and coupling
constants) of corresponding resonances. It was shown also
that the generating equations for those conditions can be
presented in the form, clearly displaying the remarkable
property which is commonly called as duality (see, for example,
\cite{6,7}). The numerical results obtained in refs.
\cite{1,2,8,9,10}
for the low energy expansions coefficients of the
corresponding amplitudes proved to be in a good agreement
with experimental data both in normal
and anomalous sectors. Altogether, these results look promising
enough and, hence, it makes sense to consider the method used in
refs. \cite{1,2,8,9,10} in more detail.

The results of the papers \cite{1,2}
have been derived in a manner strongly exploiting the high
degree of the corresponding amplitudes symmetry, the latter
one being caused solely by the symmetry of the considered
processes. Such a way, however, could not be used for the
consideration of less symmetric reactions. So, it looks
quite natural to apply Weinberg's asymptotic condition
(with the modifications suggested in ref.
\cite{11}) to a consideration of general scattering process.
This could be done in a quite general way with a help
of Weinberg's formalism \cite{12}
of Feynman rules for any spin (see also \cite{13}).

Here, however, we use another--more phenomenological--
way. This way is much more simple and transparent from the
purely technical point of view and, at the same time,
general enough to demonstrate all specific features of the
approach. We consider (as an example) three ($\pi, K$)
processes ($s-\, ,t-\,$ and $u-$ channel)
under the assumption of $SU_2$ symmetry. The amplitude of
this reaction does not possess
the high degree of symmetry inherent to the amplitudes of
$\pi\pi$ elastic scattering and
$\gamma\pi\rightarrow\pi\pi$
reaction. This gives us a possibility to show the general
way in which the final results can be derived from the
starting position. As the case of the \underline{forward}
$\pi K$ scattering is analyzed in detail in our previous paper
\cite{9},
we concentrate here mostly on the analytical aspects of the
problem rather than on numerical estimates.

Later on we use the term ``tree approximation''
instead of ``large-$N_c$ limit'' used in our early papers
\cite{1,2,8,9,10,14} .
The point is that these two terms have a quite different
meaning when applied to the case of processes with
fermions. As the approach under consideration works
equally well for both $\pi\pi$ and $\pi N$
scattering, we cannot claim that it is based on the color
number $N_c$ largeness.
It looks much as we deal with a manifestation
of some very complicated dynamical symmetry (such a
possibility has been pointed out recently by Weinberg \cite{15})
rather than with a consequence of large-$N_c$ limit.

The main goal of this paper is to attract the readers'
attention to a very important feature of every field theory
with infinite spectrum of bound states: in such a theory
even the simplest --- tree-level --- amplitude is
underdetermined. Since it takes a form of infinite sum of
pole terms, one has to define correctly the summation
procedure. There exist several ways to complete the
definition, one of them (the most natural from the purely
intuitive point of view) being analyzed below. We consider
the tree-level amplitude of a given binary scattering
process as a \underline{meromorphic}
function of 3 dependent complex variables. We take a
\underline{postulate}
that this function (the ``generalized'' amplitude) is
polynomially bounded in each energy-like variable ($s, t, u$)
at zero value of the corresponding momentum transfer.

It is shown that the above postulate unambiguously
gives rise to the following results:

1. The only parameters completely determining the amplitude
are the \underline{on-shell} couplings and masses.

2. These parameters are restricted by a certain (infinite)
system of \underline{bootstrap} equations.

3. The full cross-symmetric amplitude takes the typically
\underline{dual} form, the \underline{Pomeron}
contribution being taken into account.

4. This latter contribution corresponds to a nonresonant
\underline{background}, which, in turn, is expressed in terms of
the cross-channel resonances parameters. In other words,
it shows all properties, first suggested by Harari \cite{15a}.

5. The terms {\it  tree-level analyticity, duality} and
{\it bootstrap} are completely equivalent.

6. The bootstrap equations provide the necessary conditions
for the \underline{renormalizability}
of a theory incorporating particles of higher ($J>1$) spins.

7. Every coefficient of the Chiral Perturbation Theory (ChPT)
expansion in the chiral limit takes a form of a sum over
certain resonance contributions.

Some possible ways of further development of the
approach are discussed in Sect. 10.

\section{Preliminary notes.}
    \mbox{}

To explain better the main ingredients of our approach,
it looks useful to recall the reader some details of the
formalism describing spin-$J$ particles. Here we consider $J$
as an arbitrary integer number; the results for half-integer
values of $J$ can be found in refs. \cite{12,13}
(see also the excellent summary in \cite{16}).

The propagator of a particle with spin $J$
and nonzero mass M takes the following form:
\begin{equation} \label{2.1}
{ \cal P}^{\mu_1\cdots\mu_J}_{\nu_1\cdots\nu_J}(q,M) =
\frac{i}{(2\pi)^4}\frac{(-1)^J}{q^2-M^2}
\{ { \cal P}^{\mu_1\ldots\mu_J}_{\nu_1\ldots\nu_J}(q) +
 (OST) \}
\end{equation}
Here the abbreviation OST (Off-Shell Terms) stands for all
the terms disappearing on the mass shell $q^2 = M^2$.
The first -- explicitly written -- term in (\ref{2.1})
is the uniquely determined symmetric traceless rank-$2J$
tensor: the projecting operator on spin-$J$
states. Its explicit form, along with many other useful
formulas, can be found in ref. \cite{17}.

In what follows, however, we work with the more simple
object -- the so-called contracted projector:
\begin{equation} \label{2.2}
{ \cal P}^J (q,k,p)\equiv k_{\mu_1}\ldots k_{\mu_J}
{ \cal P}^{\mu_1\ldots\mu_J}_{\nu_1\ldots\nu_J}(q)
p^{\nu_1}\ldots p^{\nu_J}  \; .
\end{equation}
The expression (\ref{2.2}) can be rewritten as follows:
\begin{equation} \label{2.3}
{ \cal P}^J (q,k,p)
=\frac{J!}{(2J-1)!!} |{\tilde k}|^J |{\tilde p}|^J P_J (x) \; ,
\end{equation}
where $P_J$ stands for the ordinary Legendre polynomial,
\begin{equation} \label{2.4}
{\tilde k}\equiv k - \frac{k q}{M^2} q ;\;\;\;\;
{\tilde p}\equiv k - \frac{k q}{M^2} q \; ,
\end{equation}
and
\begin{equation} \label{2.5}
x\equiv\frac{{\tilde k}{\tilde p}}{|{\tilde k}| |{\tilde p}|} \; .
\end{equation}

With the help of (\ref{2.1}) --- (\ref{2.5})
one can easily construct the most general form of tree-level
amplitude of the scattering process
\begin{equation} \label{2.6}
X(k)+Y(p)\rightarrow X(k')+Y(p') \; .
\end{equation}
Such a form corresponds to an account taken of all admissible
$s-\, , t-\,$ and $u-$
channel exchanges along with the contribution of a point-like
$XXYY$-vertex. For simplicity, we consider here only the case of
spinless $X\, ,Y\,$ ; the generalization is straightforward.

Let us write down the explicit form of a contribution due to
the $s-$channel exchange of a resonance $R$ with spin $J$
and mass $M$.
To do this, one has to specify first the form of the $XYR$-vertex.
The latter one can be written as follows:
\begin{equation} \label{2.7}
V[R(q)\rightarrow X(k)Y(p)] = i (2\pi)^4\delta (q-k-p)
V^{\mu_1\cdots\mu_J}(k,p){\varepsilon}^{j}_{\mu_1\cdots\mu_J}(q)\; ,
\end{equation}
where
$\varepsilon^{j}_{\mu_1\cdots\mu_J}(q)$
stands for the wave function of spin-$J$ particle, $j$
marking the polarization. The explicit expression for the
vertex function $V^{\mu_1\cdots\mu_J}$ reads:
\begin{equation} \label{2.8}
V^{\mu_1\cdots\mu_J} (k,p) =
g^{}_{XYR}(J,M^2) k^{\mu_1}\ldots k^{\mu_J} +(OST) \; .
\end{equation}
Here, again, the abbreviation  OST stands for all the terms which
do not contribute to the RHS of eq.
(\ref{2.7}).
Such terms appear, for example, if one includes the
nonminimal couplings in the interaction Lagrangian.
The value of the coupling constant $g^{}_{XYR}$ in (\ref{2.8})
depends only on the resonance $R$ parameters (mass and spin).

Using (\ref{2.1}), (\ref{2.3}) and (\ref{2.8})
one can present the desired $S$-channel contribution
as follows:
\begin{equation} \label{2.9}
A^{(s)} = - G^{(s)}(J,M^2)
\frac{P_J (1+\frac{t}{2\Phi})}{s - M^2} + \Pi^{(s)} (s,t,u)\; ,
\end{equation}
with
\begin{equation} \label{2.10}
\Phi\equiv\Phi(M^2)=\frac{1}{4M^2}
\{ M^4 - 2M^2(m^2_x + m^2_y) + (m^2_x - m^2_y)^2 \} \; ,
\end{equation}
\begin{equation} \label{2.11}
G^{(s)}(J,M^2)\equiv |g^{}_{XYR}|^2 \frac{J!}{(2J-1)!!} |\Phi|^J \; ,
\end{equation}
and $\Pi^{(s)} (s,t,u) $
being a polynomial which contains the information both on OST in
(\ref{2.1}) and (\ref{2.8})
as well as on the detailed structure of the point-like
$XXYY$-vertex (we consider the local interactions only).

The contribution due to the $u$-channel exchange takes precisely
the same form as that given by (\ref{2.9}) with $u$
substituted instead of $s$.

At last, the $t$-channel exchange of a resonance $Z$ with spin $J$
and mass $M$ results in the following contribution:
\begin{equation} \label{2.12}
A^{(t)} = - G^{(t)}(J,M^2)
\frac{P_J (\frac{s-u}{4F})}{t - M^2} +
\Pi^{(t)} (s,t,u) \; ,
\end{equation}
where
\begin{equation} \label{2.13}
F\equiv \frac{1}{4}
\left|\sqrt{(M^2 - 4m^2_x)(M^2 - 4m^2_y)}\right|\; ,
\end{equation}
and
\begin{equation} \label{2.14}
G^{(t)}(J,M^2)\equiv g^{\ast}_{XXZ} g^{}_{YYZ}
\frac{J!}{(2J-1)!!} (F)^J \; .
\end{equation}
The value of
$G^{(s)}(J,M^2)\equiv G^{(u)}(J,M^2)$
can be connected with the decay width

\noindent $\Gamma(R\ttoo X+Y)$
(if $M>m_x+m_y$ ):
\begin{equation} \label{2.15}
G^{(s)}(J,M^2) = 8\pi M^2 (2J+1)
\frac{\Gamma (R\rightarrow X+Y)}{\sqrt{\Phi}} \; .
\end{equation}
The similar expression for
$G^{(t)}$:
\begin{equation} \label{2.16}
G^{(t)}(J,M^2) = 8\pi M^2 (2J+1)
\sqrt{ \frac{\Gamma (Z\rightarrow X\pi)
\Gamma (Z\rightarrow X\pi)}{F} }
\end{equation}
is valid under the condition
$M_Z>\max\{ 2m_x,2m_y\}$.
In the case when there are identical particles in the set
($X, \overline{X},Y, \overline{Y}$),
the Bose-factor $1/(2!)$ must be included in the corresponding
formulas.

Thus, the general tree-level amplitude $A(s,t,u)$ describing the
($X,Y$) process (\ref{2.6})
along with two cross-conjugated reactions:
\begin{eqnarray} 
X+\overline X &\ttoo & Y+\overline Y \; ,
\nonumber\\
X+\overline Y &\ttoo & X+\overline Y \; ,
\nonumber
\end{eqnarray}
can be presented in the following form:
\begin{eqnarray} \label{2.17}
A(s,t,u) &=&
 - \sum_{(XY)}^{} G^{(s)}(J,M^2) P_J \left( 1+\frac{t}{2\Phi}\right)
\left \{  \frac{1}{s-M^2} +
\frac{1}{u-M^2} \right \} \: -
\nonumber\\ &&
- \sum_{(X{\overline X})}^{} G^{(t)}(J,M^2)
\frac{P_J (\frac{s-u}{4F})}{t - M^2} +
\Pi(s,t,u) \; .
\end{eqnarray}
Here the summation is implied over
\underline{all}
resonances admissible in a given channel. The constants
$G^{(s)}$ and $G^{(t)}$ define the \underline{on-shell} couplings
$RXY$, $RX\overline X$,  $RY\overline Y$;
all details of the off-mass-shell dynamics being collected in the
polynomial (possibly, the entire  function) $\Pi(s,t,u)$.
Needless to say, the tree-level amplitude can be written in the form
(\ref{2.17}) irrelevantly to a particular dynamical language
(Lagrangian, dispersion relations, etc) used.
This is the reason why the above formalism
is called sometimes ``nondynamical'' (see, e.g. \cite{18}).

The expression (\ref{2.17}),
by itself, is well determined only in the case if the number of
admissible resonances is finite. In contrast, if the spectrum
is infinite, the form (\ref{2.17})
requires special determination in order to avoid problems connected
with a possible divergence of the summation procedure
(see Sect.4 below).
Until such a determination is specified, the eq.(\ref{2.17})
should be treated as a \underline{formal construction}.

Later on we oftenly refer to the Cauchy  method allowing one to
write down the \underline{convergent}
partial fraction expansion  of a given function $f(z)$
of  one complex variable $z$.
Inasmuch as we need only the resulting formula, it makes sense
to cite it here. Let $p_i$ ($i=1,2, \ldots$) be the poles locations
($|p_i|<|p_{i+1}|$) and $r_i$ --
the corresponding  residues (we consider here only the case of simple
poles). Next, let us specify the degree $N$
of the asymptotic grows of $f(z)$ by the condition:
\begin{equation} \label{2.18}
\int\limits_{C_n}\left|\frac{f(z)dz}{z^{N+2}}\right|
\ttoo\limits_{n\to\infty}  0 \; ,
\end{equation}
where $C_n$ is (for definiteness) a circle with the radius
$R_n$: $|p_n|<R_n<|p_{n+1}|$.
In this case the Cauchy method gives:
\begin{equation} \label{2.19}
f(z)= \sum_{n=0}^{N}f^{(n)}(0)\frac{z^n}{n!}
+ \sum_{i=1}^{\infty}
\left[ \frac{r_i}{z-p_i }- P_i (z)\right] ,
\end{equation}
where $P_i(z)$ stands for the first $N$
terms of the power series expansion of $\frac{r_i}{z-p_i}$
around the point $z=0$ ($P_i(z)$
are commonly called as ``correcting polynomials''). The convergence
of the expansion (\ref{2.19}) at any fixed value of $z$
is guaranteed by the condition (\ref{2.18}).
This is a special form of the general Mittag-Leffler theorem.

Below (Sect.8) we need also a particular  form  of (\ref{2.19}),
specially adjusted for the case when the principal part (taken alone)
converges. In this case the infinite series of correcting
polynomials, appearing in the eq. (\ref{2.19}),
converges also and, hence, it can be summed independently.
Let us restrict ourselves by a consideration of the special case of
``asymptotically constant'' function
$f(z)$ (this corresponds to $N=0$ in (\ref{2.18})).
Under the above conditions one can rewrite (\ref{2.19}) as follows:
\begin{equation} \label{2.20}
f(z)=\left \{ f(0)+\sum_{i=1}^{\infty}\frac{r_i}{p_i}
\right \}
+ \sum_{i=1}^{\infty}\frac{r_i}{z-p_i} \; .
\end{equation}

At last, we would like to note, that each coefficient
($f^{(n)}(0)$, $r_i$, $p_i$) appearing in (\ref{2.19})
can, in turn, depend on some parameter $t$;
this dependence need not be regular.

\section{General formalism for ($\pi ,K$) processes}
\mbox{}

    In what follows we consider three
($\pi ,K$)
processes:
\begin{equation} \label{3.1}
\pi_a (k_1) + K_\alpha (p_1)
\rightarrow
\pi_b (k_2) + K_\beta (p_2) \; ,
\end{equation}

\begin{equation} \label{3.2}
\pi_a (k_1) + {\overline K_\beta} (p_2)
\rightarrow
\pi_b (k_2) + {\overline K_\alpha} (p_2) \; ,
\end{equation}

\begin{equation} \label{3.3}
\pi_a (k_1) + \pi_b (k_2)
\rightarrow
{\overline K_\alpha} (p_1) + K_\beta (p_2) \; ,
\end{equation}

\noindent
Here $a,b=1,2,3$ and $\alpha ,\beta =1,2$
stand for the isotopic indices of pions and kaons, respectively.

The processes (\ref{3.1}) --- (\ref{3.3})
are connected with each other by the crossing transformation and ---
in accordance with the crossing symmetry principle --- can be
described by the unique amplitude $M^{ba}_{\beta\alpha}$:

\begin{equation} \label{3.4}
M^{ba}_{\beta\alpha}=\delta_{ba}\delta_{\beta\alpha}
A(s,t,u)+i\varepsilon_{bac}(\sigma_c)_{\beta\alpha}
B(s,t,u) \; .
\end{equation}
The isotopic amplitudes $A$ and $B$ appearing in (\ref{3.4})
depend on the standard kinematical variables

\begin{equation} \label{3.5}
s=(k_1 + p_1)^2 ; \;\; t=(k_1 - k_2)^2 ; \;\; u=(k_1 - p_2)^2 ;
\end{equation}
obeying the condition

\begin{equation} \label{3.6}
s+t+u=2(m^2 +\mu^2)\equiv 2\sigma \; ,
\end{equation}
where $m\; (\mu)$
is the kaon (pion) mass. According to the Bose symmetry requirement:

\begin{equation} \label{3.7}
A(s,t,u)=A(u,t,s) \; , \atop
B(s,t,u)=-B(u,t,s) \; .
\end{equation}

Sometimes, it is convenient to use the set $(\nu ,t)$
of two independent variables, with
\begin{equation} \label{3.8}
\nu\equiv s-u \; ,
\end{equation}
the corresponding expressions for $s$ and $u$ being of the form:

\begin{equation} \label{3.9}
s={\nu\over 2} + {2\sigma-t\over 2}\; , \;\;
u=-{\nu\over 2} + {2\sigma-t\over 2}\; ,
\end{equation}

Later on we take \underline{a postulate}
that no exotic mesons exist. So, in the case under consideration
only nonstrange mesons with the isospin
$I=0,1$ (and positive normality) along with strange ones with
$I=1/2$ contribute. The resulting tree-level amplitudes
$A$ and $B$ can be written as follows (see Sec.2):
\begin{displaymath}
A(s,t,u)= -
\sum_{(I=0)} G_0
\frac{P_J \left(\frac{s-u}{4F}\right)}{t-M^2} \; -
\sum_{(I=1/2)}G_{1/2}
P_J\left(1+{t\over 2\phi}\right)
\left\{\frac{1}{s-M^2} +\frac{1}{u-M^2}
\right\}  +
\end{displaymath}
\begin{displaymath}
 + \Pi_A (s,t,u) \; ,
\end{displaymath}
\begin{displaymath}
B(s,t,u)= -
\sum_{(I=1)} G_1
\frac{P_J \left(\frac{s-u}{4F}\right)}{t-M^2}\; -
\sum_{(I=1/2)} G_{1/2}
P_J\left(1+{t\over 2\phi}\right)
\left\{\frac{1}{s-M^2} -\frac{1}{u-M^2}
\right\}  +
\end{displaymath}
\begin{equation} \label{3.10}
+ \Pi_B (s,t,u) \; ,
\end{equation}
where (due to (\ref{3.7}))
\begin{equation} \label{3.11}
\Pi_A (s,t,u)=\Pi_A(u,t,s)\; ,
\atop
\Pi_B (s,t,u)=-\Pi_B(u,t,s)\;
\end{equation}

Each symbol $\sum\limits_{(I=p)}$ appearing in (\ref{3.10})
implies a summation over \underline{all}
admissible resonances with the indicated $(p=0,1/2,1)$
isospin value. For example:
\begin{equation} \label{3.12}
\sum_{(I=0)} G_0
\frac{P_J \left(\frac{s-u}{4F}\right)}{t-M^2}
\equiv
\sum_{i=1}^{\infty}\frac{R_i(M_i , \nu)}{t-{M_i}^2} \; ,
\end{equation}
where ${M_i}^2 < {M_{i+1}}^2$ and
\begin{equation} \label{3.13}
R_i \equiv
\sum_{J=0,2,\cdots} G_0 (J,{M_i}^2)
P_J \left(\frac{\nu}{4F(M_i)}\right) \; .
\end{equation}

It should be particularly emphasized, that the outlined above
summation order is very important: the inner sum has to be taken
over all resonances with a given mass $M_i$ ,
the outer one --- over the mass values
\underline{in order of increasing mass}.
This (and only this) manner of summation makes it possible to
consider the forms like (\ref{3.10})
as partial fraction expansions in a space of 3 complex variables
$s,t,u$ .

In other words, we take the inner sum (\ref{3.13}) \underline{across}
the Regge (or, more precisely, Khuri) trajectories and, hence, take
account of all possible satellites. We \underline{assume}
that this procedure produces the \underline{finite} residues
$R_i ({M_i}^2 ,\nu)$ to make it sensible the outer sum (\ref{3.12}) :

\begin{equation} \label{3.14}
R_i ({M_i}^2 , \nu) \leq Q_i (\nu) < \infty \; ,\;\;
(i=1,2,\ldots) \; .
\end{equation}
The requirement (\ref{3.14})
is, in fact, unnecessarily strong: as it is shown below, it turns
out sufficient
to require the finiteness of residues at one fixed value of the
corresponding kinematical variable. This assumption, in turn, implies
certain limitations on the coupling constants $G_i(J,{M_i}^2)$
dependence of their arguments: the existence of the leading
trajectory would be quite enough. Later on, however, we do not use
any particular form of those limitations.

\section{Asymptotic condition.}
\mbox{}

The given above general form (\ref{3.10}) of the tree-level amplitude
(\ref{3.4}) describing three ($\pi ,K$) processes
(\ref{3.1})---(\ref{3.3}) contains two unspecified functions
$\Pi_A$, $\Pi_B$.
To fix them, we attract the generalized version of Weinberg's
asymptotic requirement \cite{3} (see also \cite{15}).
The original formulation is changed in two points. First, we do
not require pion to be massless \cite{11}.
Second, the asymptotic requirement is thought to be suitable for
\underline{every}
binary scattering process at zero value of the momentum transfer.

So, our formulation of the asymptotic condition reads:
{\em the high energy behavior of the tree-level amplitude of a
given binary scattering process at zero momentum transfer must
not violate the experimentally known boundary}.

In other words, we suppose that --- due to some special mechanism
(a kind of dynamical symmetry?) --- all rapidly increasing with
energy tree-level contributions cancel among themselves at zero
value of the momentum transfer. It should be specially stressed,
that we \underline{do not require}
the true experimental (or, the same, Regge) behavior of the
tree-level amplitude: we require only the polynomial boundedness,
the corresponding degree being dictated by the experiment. As to
the inelastic processes amplitudes, we require of them merely to
decrease with energy. So, our requirements are, in a sense, weak
enough.

At the same time, Weinberg's results (as well as those derived in
refs. \cite{24,25,21,22,29,26,27,23,28,1,2,8,9,10,14,19,20})
clearly demonstrate that it is feasible to analyze the asymptotic
behavior degree by degree. Such a way could be only applied if
the analytical structure of the amplitude is simple enough. This
note shows that it makes sense to consider the problem in question
in terms of analytical functions from the very beginning.

In the opening stage of the following analysis we use the more
strong formulation of the asymptotic condition: we require the
``realistic'' asymptotic behavior of the amplitude not only at
$t=0$, but also at arbitrary nonpositive $t$
from a small neighborhood of the point $t=0$.
This is done purely for the sake of reader's convenience. As we
prove below (see Sect.7), Weinberg's formulation (requiring $t=0$
only) appears to be quite sufficient to provide the correctness
of our results.

To proceed further, we have to specifY the experimental (or, the
same, given by Regge-like fits) limitations for the amplitudes
$A(s,t,u)$ and $B(s,t,u)$ appearing in (\ref{3.4}).

Let us consider first the inelastic process (\ref{3.3}).
In this case both amplitudes $A$ and $B$
must decrease with energy (i.e. at $t \to \infty$).
Since there is a freedom to choose either $s$ or $u$
for a momentum transfer, we can write down two sets of the
asymptotic conditions:
\begin{equation} \label{4.1}
\left\{
\begin{array}{lll}
A(s,t,u)\ttoo\limits_{t\to\infty} 0 \; ,  \\
B(s,t,u)\ttoo\limits_{t\to\infty} 0 \; ,  \\
(s,t,u)\in {\cal D}_{ts} \; ,
\end{array}
\right.
\end{equation}
and
\begin{equation} \label{4.2}
\left\{
\begin{array}{lll}
A(s,t,u)\ttoo\limits_{t\to\infty} 0 \; , \\
B(s,t,u)\ttoo\limits_{t\to\infty} 0 \; , \\
(s,t,u)\in {\cal D}_{tu} \; .
\end{array}
\right.
\end{equation}
Here we have used the symbol ${\cal D}_{xy}$
to denote the band where real $x$ plays a role of the CMS
energy squared, while small nonpositive $y$ ---
of the momentum transfer. The meaning of the term ``small'' is
explained below (Sect.7).

Next, let us consider two elastic processes (\ref{3.1}) and
(\ref{3.2}).
In both cases the asymptotic behavior is governed by the
Pomeron for $A$ ($\alpha_0 (0) = 1$) and by the
$\rho$-trajectory for $B$ ($\alpha_1 (0) \approx 0.5$). So, we have:
\begin{equation} \label{4.3}
\left\{
\begin{array}{lll}
s^{-2}A(s,t,u)\ttoo\limits_{s\to\infty} 0 \; ,\\
s^{-1}B(s,t,u)\ttoo\limits_{s\to\infty} 0 \; ,\\
(s,t,u)\in {\cal D}_{st} \; ,
\end{array}
\right.
\end{equation}
and
\begin{equation} \label{4.4}
\left\{
\begin{array}{lll}
u^{-2}A(s,t,u)\ttoo\limits_{u\to\infty} 0 \; ,\\
u^{-1}B(s,t,u)\ttoo\limits_{u\to\infty} 0 \; ,\\
(s,t,u)\in {\cal D}_{ut} \; ,
\end{array}
\right.
\end{equation}

It is clear that, with the help of eq. (\ref{3.7}), two ``elastic''
conditions (\ref{4.3}) and (\ref{4.4})
could be rewritten in a unique (more strong) form. This point
will be discussed later.

In what follows we assume that there exist two functions
$\overline{A}$ and $\overline{B}$ (we call them as the
``generalized'' amplitudes) of three dependent \underline{complex}
variables $s, t, u$ (or, two independent ones: $\nu , t$),
each of them satisfying the following conditions:
\begin{description}
\item{A.}
At real $s, t, u$
it coincides identically with the corresponding physical amplitude
\mbox{($A$ or $B$).}
\item{B.}
It is a meromorphic function with no other poles but those
explicitly shown in \mbox{eqs.(\ref{3.10}).}
\item{C.}
When being considered as a function of one \underline{complex}
variable $\overline{x}$
(the CMS energy squared in a given channel) and one real
(small, nonpositive) parameter $y$
(the momentum transfer), it is polynomially bounded in
$\overline{x}$,
the degree of corresponding polynomials being dictated by
the asymptotic conditions (\ref{4.1}) ---(\ref{4.4}).
\end{description}

As it is mentioned above, the requirement $y \leq 0 $
appearing in the point C, can be reduced to $y = 0 $.

Now, we are in a position to derive the results following from
the formulated above requirements.

\section{Elastic ($\pi , K$) processes.}
\mbox{}

Let us begin from the detailed consideration of the elastic $\pi K$
-- scattering process (\ref{3.1}).
To apply the asymptotic condition one has to consider the generalized
amplitudes ${\overline A}$ and ${\overline B}$ in the area
${\cal D}_{\overline{s} t}$ of arbitrary complex $s$
and real (small, nonpositive) values of $t$
(hereafter we use --- if necessary --- the symbol $\overline{x}$
to denote the complex-valued variable $x$).

In accordance with the assumption B (Sect.4) and the eq.(\ref{3.10}),
at every fixed $t\in {\cal D}_{\overline{s}t}$ both amplitudes
$\overline{A}$ and $\overline{B}$
have only simple poles, the latter ones being located on the axis
$\mbox{Im}\:s = 0$ in the following points:
\begin{equation} \label{5.1}
\left\{
\begin{array}{lll}
s_i={M_i}^2 \; , & \mbox{(fixed poles)}\; , \\
s_i=-(\Sigma_i + t)\; , & \mbox{(moving poles)}\; ,\\
\Sigma_i\equiv {M_i}^2 - 2\sigma, & \mbox{(i=1,2,\ldots).}
\end{array}
\right.
\end{equation}
Next, according to the eq. (\ref{4.3})
and the assumption C, these functions obey in
${\cal D}_{\overline{s}t}$ the following boundedness conditions (see
(\ref{2.18}) for the notation $C_n$):
\begin{displaymath}
\int\limits_{C_n}
\left|
\frac{{\overline A}({\overline s},t)}{{\overline s}^3} d{\overline s}
\right|
\ttoo\limits_{n\to\infty} 0\; ,
\end{displaymath}
\begin{equation} \label{5.2}
\int\limits_{C_n}
\left|
\frac{{\overline B}({\overline s},t)}{{\overline s}^2} d{\overline s}
\right|
\ttoo\limits_{n\to\infty} 0\; .
\end{equation}
Hence, applying the Cauchy formula (\ref{2.19}) (with the eq.
(\ref{3.10}) taken into account), one obtains the following
\underline{convergent} partial fraction expansions for the amplitudes
$\overline{A}$ and $\overline{B}$ in ${\cal D}_{\overline{s}t}$:
\begin{displaymath}
{\overline A}(s,t,u)=
{\overline A}(t_s)+
s\frac{\partial {\overline A}(t_s)}{\partial s} \; -
\sum_{(I=1/2)} G_{1/2} P_J\left( 1+{t\over 2\Phi} \right)
\left\{
\left(\frac{1}{s-M^2}+\frac{1}{u-M^2} \right)+
\right.
\end{displaymath}
\begin{displaymath}
\left.
+ \left(\frac{1}{M^2}+\frac{1}{\Sigma +t} \right)+
s \left(\frac{1}{M^4}-\frac{1}{(\Sigma +t)^2} \right)
\right\}\; ,
\end{displaymath}
\begin{displaymath}
{\overline B}(s,t,u)=
{\overline B}(t_s)-
\end{displaymath}
\begin{equation} \label{5.3}
-\sum_{(I=1/2)} G_{1/2} P_J\left( 1+{t\over 2\Phi} \right)
\left\{
\left(\frac{1}{s-M^2}-\frac{1}{u-M^2} \right)+
\left(\frac{1}{M^2}-\frac{1}{\Sigma +t} \right)
\right\} \; .
\end{equation}
Here
\begin{equation} \label{5.4}
(t_s)\equiv (0,t,2\sigma - t)
\end{equation}
and $u=2\sigma -s-t \; . $

We would like to stress again that --- in contrast with the
formally written expressions (\ref{3.10}) ---
both partial fraction expansions (\ref{5.3}) are convergent in
${\cal D}_{\overline{s}t}$ \underline{by construction}
based on the postulated above asymptotic condition: the
convergence is guaranteed by the Mittag-Leffler theorem. This
very theorem dictates one to include the correcting polynomials
in $s$ into each item of the sums in (\ref{5.3}),
their minimal degree being uniquely connected with the asymptotic
behavior. At the same time, as it is clear from the given above
formulas (\ref{5.3}),
the coefficients of those polynomials contain unwanted singularities
--- the fixed poles in $t$
of the first and second orders, these poles appearing at
\underline{negative} $t_i = -{\Sigma}_i$.
The problem of the second order poles is solved in the end of this
Section; the detailed discussion of the problem of first order
poles at negative values of $t$
can be found in Sections 7 and 8 below.

The other two essential features differing (\ref{5.3}) from
(\ref{3.10})
are the following. First, instead of two unknown functions of
$s$ and $t$ ($\Pi_A$ and $\Pi_B$) appearing in (\ref{3.10}),
one has in (\ref{5.3}) three unspecified functions of $t$ only
(${\overline A}(t_s), {\overline B}(t_s)$ and
$\frac{\partial {\overline A}(t_s)}{\partial s}$).
Second, the Bose and crossing symmetries, clearly visible in
(\ref{3.10}), seem to be completely lost in (\ref{5.3}).
We shall come back to this problem somewhat below.

Let us now turn to a consideration of the second elastic $(\pi ,K)$
process (\ref{3.2}). In the area ${\cal D}_{\overline{u}t}$
(arbitrary complex $u$, small nonpositive $t$)
the given above argumentation can be repeated word by word.
This allows one to write down the corresponding partial fraction
expansion immediately. So, in every point
$(s,t,u)\in {\cal D}_{\overline{u}t}$ we have:
\begin{displaymath}
{\overline A}(s,t,u)=
{\overline A}(t_u)+
u \frac{\partial {\overline A}(t_u)}{\partial u} \; -
\sum_{(I=1/2)} G_1/2 P_J\left( 1+{t\over 2\Phi} \right)
\left\{
\left(\frac{1}{s-M^2}+\frac{1}{u-M^2} \right)+
\right.
\end{displaymath}
\begin{displaymath}
\left.
+ \left(\frac{1}{M^2}+\frac{1}{\Sigma +t} \right)+
u \left(\frac{1}{M^4}-\frac{1}{(\Sigma +t)^2} \right)
\right\}\; ,
\end{displaymath}
\begin{displaymath}
{\overline B}(s,t,{\overline u})
={\overline B}(t_u)-
\end{displaymath}
\begin{equation} \label{5.5}
-\sum_{(I=1/2)} G_1/2 P_J\left( 1+{t\over 2\Phi} \right)
\left\{
\left(\frac{1}{s-M^2}-\frac{1}{u-M^2} \right)-
\left(\frac{1}{M^2}-\frac{1}{\Sigma +t} \right)
\right\} \; .
\end{equation}
where
\begin{equation} \label{5.6}
(t_u)\equiv (2\sigma - t,t,0)\; ,
\end{equation}
and the symbol ${\Sigma}_i$ is determined in (\ref{5.1}).
Comparing to (\ref{5.3}), three new unspecified functions ---
${\overline A}(t_u), {\overline B}(t_u)$ and
$\frac{\partial {\overline A}(t_u)}{\partial u}$ --- appear in
(\ref{5.5}).

The important note, giving one a key to the subsequent progress,
can now be formulated in two steps.
\begin{enumerate}
\item
The intersection area ${\cal D}_{{\overline \nu}t}$:
\begin{displaymath}
{\cal D}_{{\overline \nu}t}={\cal D}_{{\overline s}t}\cap
{\cal D}_{{\overline u}t}
\end{displaymath}
is nonempty.
\item
Since both forms (\ref{5.3}) and (\ref{5.5}) determine
\underline{the same} analytical function ${\overline A}(s,t,u)\;$
(${\overline B}(s,t,u)$) , they must coincide identically in
${\cal D}_{{\overline \nu}t}$ (arbitrary complex $\nu$,
small nonpositive $t$).
\end{enumerate}

Thus, two equalities are hold in ${\cal D}_{{\overline \nu}t}$:
\begin{displaymath}
{\overline A}(t_s)+
s\frac{\partial {\overline A}(t_s)}{\partial s} -
\sum_{(I=1/2)} G_{1/2} P_J\left( 1+{t\over 2\Phi} \right)
\left\{
\left(\frac{1}{s-M^2}+\frac{1}{u-M^2} \right)+
\right.
\end{displaymath}
\begin{displaymath}
\left.
+ \left(\frac{1}{M^2}+\frac{1}{\Sigma +t} \right)+
s \left(\frac{1}{M^4}-\frac{1}{(\Sigma +t)^2} \right)
\right\}\; =
\end{displaymath}
\begin{displaymath}
=\; {\overline A}(t_u)+
u \frac{\partial {\overline A}(t_u)}{\partial u}\; -
\sum_{(I=1/2)} G_{1/2} P_J\left( 1+{t\over 2\Phi} \right)
\left\{
\left(\frac{1}{s-M^2}+\frac{1}{u-M^2} \right)+
\right.
\end{displaymath}
\begin{displaymath}
\left.
+ \left(\frac{1}{M^2}+\frac{1}{\Sigma +t} \right)+
u \left(\frac{1}{M^4}-\frac{1}{(\Sigma +t)^2} \right)
\right\}\; ,
\end{displaymath}
\begin{displaymath}
{\overline B}(t_s)
-\sum_{(I=1/2)} G_{1/2} P_J\left( 1+{t\over 2\Phi} \right)
\left\{
\left(\frac{1}{s-M^2}-\frac{1}{u-M^2} \right)+
\left(\frac{1}{M^2}-\frac{1}{\Sigma +t} \right)
\right\} \; =
\end{displaymath}
\begin{equation} \label{5.7}
={\overline B}(t_u)
-\sum_{(I=1/2)} G_{1/2} P_J\left( 1+{t\over 2\Phi} \right)
\left\{
\left(\frac{1}{s-M^2}-\frac{1}{u-M^2} \right)-
\left(\frac{1}{M^2}-\frac{1}{\Sigma +t} \right)
\right\} \; .
\end{equation}
Separating the independent variables $(\nu , t)$ in the first of eqs.
(\ref{5.7}) with the help of (\ref{3.9}), one obtains:
\begin{displaymath}
{\overline A}(t_s) - {\overline A}(t_u) + \frac{2\sigma -t}{2}
\left\{
\frac{\partial{\overline A}(t_s)}{\partial s} -
\frac{\partial{\overline A}(t_u)}{\partial u}
\right\} =0\; ,
\end{displaymath}
\begin{displaymath}
\frac{\partial{\overline A}(t_s)}{\partial s} -
\frac{\partial{\overline A}(t_u)}{\partial u} =
2\sum_{I=1/2} G_{1/2} P_J \left(1+{t\over 2\Phi} \right)
\left\{ {1\over M^4} - \frac{1}{(\Sigma + t)^2} \right\}\; ;
\end{displaymath}
\begin{equation} \label{5.8}
{\overline B}(t_s) - {\overline B}(t_u) =
2\sum_{I=1/2} G_{1/2} P_J \left(1+{t\over 2\Phi} \right)
\left\{ {1\over M^2} - \frac{1}{\Sigma + t} \right\}\; .
\end{equation}

The system of eqs. (\ref{5.8})
gives one the first series of the necessary
{\em self-consistency conditions}
reflecting the crossing symmetry requirements. They
restrict the structure of the generalized amplitudes
$\overline{A}$ and $\overline{B}$.

With the help of (\ref{5.8}) one can rewrite
(\ref{5.3}) and (\ref{5.5}) in the unique form:
\begin{displaymath}
{\overline A}(s,t,u)=
{1\over 2}\left\{ {\overline A}(t_s)+{\overline A}(t_u)\right\}
+{\nu\over 4}
\left\{\frac{\partial {\overline A}(t_s)}{\partial s} -
\frac{\partial{\overline A}(t_u)}{\partial u} \right\} -
\end{displaymath}
\begin{displaymath}
-\sum_{I=1/2} G_{1/2} P_J \left(1+{t\over 2\Phi} \right)
\left\{
\left( {1\over s-M^2} + \frac{1}{(u-M^2)} \right) +
\left( {1\over M^2} + \frac{1}{\Sigma + t} \right)-
\right.
\end{displaymath}
\begin{displaymath}
\left.
-\frac{2\sigma -t}{2}
\left( {1\over M^4} - \frac{1}{(\Sigma + t)^2} \right)
\right\}\; ,
\end{displaymath}
\begin{equation} \label{5.9}
{\overline B}(s,t,u)=
{1\over 2}\left\{ {\overline B}(t_s)+{\overline B}(t_u)\right\}
-\sum_{I=1/2} G_{1/2} P_J \left(1+{t\over 2\Phi} \right)
\left\{  \frac{1}{s-M^2} - \frac{1}{u-M^2} \right\}\; ,
\end{equation}
which is convenient for the analysis of the Bose symmetry
requirements (\ref{3.7}).
>From the comparison of (\ref{5.9}) with (\ref{3.7})
one obtains the second series of the necessary consistency
conditions:
\begin{displaymath}
\frac{\partial {\overline A}(t_s)}{\partial s} -
\frac{\partial{\overline A}(t_u)}{\partial u} =0\; ,
\end{displaymath}
\begin{equation} \label{5.10}
{\overline B}(t_s) - {\overline B}(t_u) = 0\; .
\end{equation}

The Bose symmetry requirements of the type (\ref{5.10})
play a special role because they reduce the influence of the
Pomeron contribution on the asymptotic behavior of every
meson-meson and meson-baryon elastic scattering process.

Combining (\ref{5.8}) and (\ref{5.10}), one obtains:
\begin{displaymath}
{\overline A}(t_s)={\overline A}(t_u)\equiv a(t) \; ,
\end{displaymath}
\begin{displaymath}
{\overline B}(t_s)=-{\overline B}(t_u) =
-\sum_{I=1/2} G_{1/2} P_J \left(1+{t\over 2\Phi} \right)
\left(  \frac{1}{M^2} - \frac{1}{\Sigma +t} \right)\; ,
\end{displaymath}
\begin{equation} \label{5.11}
\frac{\partial {\overline A}(t_s)}{\partial s} =
\frac{\partial{\overline A}(t_u)}{\partial u} =
-\sum_{I=1/2} G_{1/2} P_J \left(1+{t\over 2\Phi} \right)
\left(  \frac{1}{M^4} - \frac{1}{(\Sigma +t)^2} \right)\; .
\end{equation}
With the help of (\ref{5.11}) and (\ref{5.9})
one can write down the final expressions for the generalized
amplitudes $\overline{A}$ and $\overline{B}$ in
${\cal D}_{{\overline \nu}t}$:
\begin{displaymath}
{\overline A}(s,t,u)= a(t)\; -
\sum_{I=1/2} G_{1/2} P_J \left(1+{t\over 2\Phi} \right)
\left\{
\left(  \frac{1}{s-M^2} + \frac{1}{u-M^2} \right) +
\left(  \frac{1}{M^2} + \frac{1}{\Sigma + t} \right)
\right\}\; ,
\end{displaymath}
\begin{equation} \label{5.12}
{\overline B}(s,t,u)=
-\sum_{I=1/2} G_{1/2} P_J \left(1+{t\over 2\Phi} \right)
\left\{
\frac{1}{s-M^2} - \frac{1}{u-M^2}
\right\}\; .
\end{equation}
Here $a(t)$ is the only function (of one variable $t$)
which still remains unspecified in terms of the spectrum
parameters $G_I(J,M^2_i)$ and $M^2_i$.

It is needless to say that the form (\ref{5.12})
could be derived in a shorter way, if we work from the very
beginning with the Bose-symmetric asymptotic conditions written
in terms of $(\nu ,t)$.
The latter way, however, looks less instructive in comparison
with that used above. In particular, in terms of $(\nu ,t)$
it would be impossible to follow the effect of the mutual
cancellation of the second order fixed poles in $t$, appearing in
(\ref{5.3}) and (\ref{5.5}) at $t=-{\Sigma}_i$.

\section{Inelastic process $\pi\pi\to K{\overline K}$.}
\mbox{}

Let us begin the analysis of the inelastic process (\ref{3.3})
from a consideration of the general amplitudes
$\overline A$ and $\overline B$ in the area
${\cal D}_{{\overline t}s}$.
In accordance with the asymptotic conditions (\ref{4.1}),
the corresponding partial fraction expansions can be
written as follows:
\begin{displaymath}
{\overline A}(s,t,u)=
-\sum_{I=0} G_0 \frac{P_J\left(\frac{\Sigma + 2s}{4F}\right)}{t-M^2}
-\sum_{I=1/2} G_{1/2}
\frac{P_J\left(1-\frac{\Sigma + s}{2\Phi}\right)}{u-M^2} \; ,
\end{displaymath}
\begin{equation} \label{6.1}
{\overline B}(s,t,u)=
\sum_{I=1} G_1 \frac{P_J\left(\frac{\Sigma + 2s}{4F}\right)}{t-M^2}
+\sum_{I=1/2} G_{1/2}
\frac{P_J\left(1-\frac{\Sigma + s}{2\Phi}\right)}{u-M^2} \; .
\end{equation}

Indeed, the consideration of the form (\ref{3.10}) shows that in
${\cal D}_{{\overline t}s}$ each of the amplitudes
${\overline A}\;,{\overline B}$ has two sets of poles in $t$:
\begin{equation} \label{6.2}
\left\{
\begin{array}{lll}
t_i = {M_i}^2 & \mbox{(fixed poles) ,} \\
t_i = -(\Sigma_i + s) & \mbox{(moving poles) ,} \\
i=1,2,\ldots \; .
\end{array}
\right.
\end{equation}
Thus, the pole structure of ${\overline A}\;({\overline B})$ in
${\cal D}_{{\overline t}s}$ is qualitatively similar to that in
${\cal D}_{{\overline{\nu}}t}$.
At the same time, the asymptotic condition (\ref{4.1}) ---
in contrast with (\ref{4.3}) and (\ref{4.4})---
shows, that in the case under consideration there is no necessity
to include any regular terms (analogous to $a(t)$ in (\ref{5.12}))
as well as the correcting polynomials. So, the partial fraction
expansions (\ref{6.1})
are convergent by construction, their particular forms being
correlated with the asymptotic conditions (\ref{4.1}).

The only explanation is required in connection with the
summation order in (\ref{6.1}).
As it has been already pointed out in Sect.3, the form (\ref{6.1})
should be understood as a single sum over the various contributions,
the summation being implied to be done in order of increasing
modulo of pole locations (irrelevantly to the isospin values). This
very order of summation is meant throughout the paper.

Let us consider now the same process (\ref{3.3}) in the area
${\cal D}_{{\overline t}u}$.
Taking into account the asymptotic condition (\ref{4.2}),
we can repeat step by step the given above argumentation to obtain:
\begin{displaymath}
{\overline A}(s,t,u)=
-\sum_{I=0} G_0 \frac{P_J\left(\frac{\Sigma + 2u}{4F}\right)}{t-M^2}
-\sum_{I=1/2} G_{1/2}
\frac{P_J\left(1-\frac{\Sigma + u}{2\Phi}\right)}{s-M^2} \; ,
\end{displaymath}
\begin{equation} \label{6.3}
{\overline B}(s,t,u)=
-\sum_{I=1} G_1 \frac{P_J\left(\frac{\Sigma + 2u}{4F}\right)}{t-M^2}
-\sum_{I=1/2} G_{1/2}
\frac{P_J\left(1-\frac{\Sigma + u}{2\Phi}\right)}{s-M^2} \; .
\end{equation}
where $(s,t,u)\in {\cal D}_{{\overline t}u}$.

In the area ${\cal D}_t(small\; s,u\leq 0)$:
\begin{equation} \label{6.4}
{\cal D}_t \equiv {\cal D}_{{\overline t} s} \cap
{\cal D}_{{\overline t} u}\; ,
\end{equation}
both forms (\ref{6.1}) and (\ref{6.3})
are equally applicable. So, we conclude, that at
$(s,t,u)\in {\cal D}_t$:
\begin{displaymath}
\sum_{I=0} G_0 \frac{P_J\left(\frac{\Sigma + 2s}{4F}\right)}{t-M^2}
+\sum_{I=1/2} G_{1/2}
\frac{P_J\left(1-\frac{\Sigma + s}{2\Phi}\right)}{u-M^2} \; =
\end{displaymath}
\begin{displaymath}
=\; \sum_{I=0} G_0
\frac{P_J\left(\frac{\Sigma + 2u}{4F}\right)}{t-M^2}
+\sum_{I=1/2} G_{1/2}
\frac{P_J\left(1-\frac{\Sigma + u}{2\Phi}\right)}{s-M^2} \; ,
\end{displaymath}
\begin{displaymath}
\sum_{I=1} G_1 \frac{P_J\left(\frac{\Sigma + 2s}{4F}\right)}{t-M^2}
+\sum_{I=1/2} G_{1/2}
\frac{P_J\left(1-\frac{\Sigma + s}{2\Phi}\right)}{u-M^2} \; =
\end{displaymath}
\begin{equation} \label{6.5}
-\sum_{I=1} G_1 \frac{P_J\left(\frac{\Sigma + 2u}{4F}\right)}{t-M^2}
-\sum_{I=1/2} G_{1/2}
\frac{P_J\left(1-\frac{\Sigma + u}{2\Phi}\right)}{s-M^2} \; .
\end{equation}

Two relations (\ref{6.5})
give us the third series of the self-consistency conditions, the
latter ones expressing --- at the same time --- the Bose symmetry
requirements for the generalized amplitudes
$\overline A$ and $\overline B$ in ${\cal D}_t$.

The conditions (\ref{6.5}) --- in contrast with (\ref{5.11}) ---
strongly restrict the values of the spectrum parameters (masses and
coupling constants). This statement becomes evident if one considers
(\ref{6.5}) as a kind of generating equalities in ${\cal D}_t$.
Indeed, expanding both sides of each of the eqs. (\ref{6.5})
in a double series of $(\nu - {\nu}_0 \; ,t - t_0)$ around the
\underline{arbitrary} point $M_0 ({\nu}_0, t_0) \in {\cal D}_t$,
one can obtain two infinite sets of sum rules for the
parameters of resonances. The arbitrariness of $M_0$
reflects a presence of the additional --- extremely high ---
degree of the underlying symmetry.

\section{Brief digression: what means ``small'' ?}
\mbox{}

Now it is pertinent to elucidate the precise meaning of the term
``small''
used above to describe the widths of various applicability bands
${\cal D}_{{\overline x}y}$ .
>From the above consideration it is clear that the correctness of
the results is only guaranteed if this term can be changed for the
term ``finite''. Below we prove that the given above formulation of
the asymptotic condition (Sect.4) leads to a desirable finiteness
of the applicability bands. For definiteness, we consider in detail
the elastic amplitude $\overline{A}$;
the other cases can be analyzed by analogy with this one.

\underline{By construction},
based on the postulated asymptotic condition (\ref{5.2})
and the Cauchy formula (\ref{2.19}), the partial fraction expansion
(\ref{5.3}) converges at $t=0$ everywhere in the complex-$s$
plane, except, of course, the poles given by the eq.
(\ref{5.1}). What happens if we take very small (finite!) negative
$t$?
First, the second (moving) series of poles slightly moves to the
right. Second, the corresponding residues values change a little.
Third, the values of ${\overline A}(t_s)$ and
$\frac{\partial {\overline A}(t_s)}{\partial s}$
change also. This latter effect, however, cannot change considerably
the asymptotic behavior of the regular term, since, in accordance
with the assumption B (Sect.4), both ${\overline A}(t_s)$ and
$\frac{\partial {\overline A}(t_s)}{\partial s}$
are smooth (almost everywhere) functions of $t$.
So, the only question to be answered, is that of the resulting series
convergence.

To answer this question, let us consider the auxiliary partial
fraction expansion constructed precisely in accordance with
(\ref{5.3}) at $t\neq 0$, except for the pole locations (\ref{5.1}),
which have to be taken at $t=0.$ For $-4\Phi (M_1) \leq t \leq 0$:

\begin{equation} \label{7.1}
\left| P_J\left(1+\frac{t}{2\Phi
(M_i)}\right)\right|\, \leq\, 1\; ,\; \;(i=1,2,\ldots),
\end{equation}
and each term of the auxiliary expansion is majorized by the
corresponding term of (\ref{5.3}) taken at $t=0$. The expansion
(\ref{5.3}) at $t=0$
is convergent in accordance with the asymptotic condition. So, our
auxiliary expansion also converges. Now it is clear, that one can
shift the poles to their correct positions (dictated by (\ref{5.1})
at $t \neq 0$)
without breaking the convergence, the asymptotic condition
(\ref{5.2}) remaining also unchanged.

Notice, that there is no necessity to require the fulfillment of
(\ref{7.1}) for each $i=1,2,\ldots$.
To ensure the convergence, it is enough if this relation is
satisfied for all $i>i_0$.
This note, in fact, shows that the principal part of the partial
fraction expansion (\ref{5.3})
converges at arbitrary nonpositive value of the momentum transfer.
This argumentation is also applied to the expansions
(\ref{6.1}) and (\ref{6.3}).

So, we conclude, that Weinberg's requirement of the asymptotic
boundedness at zero momentum transfer appears to be quite sufficient
to guarantee the convergence of our partial fraction expansions at
\underline{arbitrary} nonpositive value of the momentum transfer.
Hence, the term ``small'' can be changed for ``finite''.

One further question, which we would like to discuss in this
Section, concerns with the behavior of the generalized amplitude
$\overline{A}$ , given by (\ref{5.12}), near the points
$t_i=-{\Sigma}_i$.
At first sight, these points correspond to a set of fixed poles in
$t$, the appearance of such poles at \underline{negative} $t$
being in contradiction with our assumption B (Sect.4). At the same
time, the corresponding terms in (\ref{5.12})
stem from the Cauchy formula (\ref{2.19});
they are necessary to ensure the convergence of the partial fraction
expansion under the conditions of polynomial boundedness (\ref{5.2})
along with the crossing and Bose symmetry. So, the problem looks
serious.

Fortunately, it is nothing but a mirage. As we show in the next
Section, these false poles are contracted by the corresponding terms
originating from the regular (in $s$!) part $a(t)$.
The similar effect has been found already above (compare
(\ref{5.3}) with (\ref{5.12})):
the complete contraction of the undesirable \underline{second order}
fixed poles in $t$
occurred as a direct consequence of the Bose symmetry requirement.

>From the above analysis it also follows that the main qualitative
effect which can (and does) occur at $t \to \infty$
reduces to a softening of the asymptotic behavior of the amplitude.
It is conditioned by an increase of the relative density of poles
in the central area of a complex-$\nu$ plane, accompanied by a
decrease of residues magnitudes. It would be of great interest to
study this effect in more detail, because it can give us a key to a
deeper understanding of the Pomeron contribution. This will be done
elsever.

\section{Bootstrap and duality.}
\mbox{}

Let us now derive the fourth (and the last) series of the
consistency conditions. This series follows from the comparison of
(\ref{5.12}) with (\ref{6.1}) in the area ${\cal D}_u$:
\begin{displaymath}
{\cal D}_u \equiv {\cal D}_{{\overline s} t} \cap
{\cal D}_{{\overline t} s}\; ,
\end{displaymath}
where both forms can be equally applied. This gives:

\begin{displaymath}
a(t)-\sum_{I=1/2} G_{1/2} P_J\left(1+\frac{t}{2\Phi}\right)
\left\{\left(\frac{1}{s-M^2}+\frac{1}{u-M^2}\right)+
\left(\frac{1}{M^2}+\frac{1}{\Sigma + t}\right)
\right\} \; =
\end{displaymath}
\begin{displaymath}
=\; -\sum_{I=0} G_0
\frac{P_J\left(\frac{\Sigma + 2s}{4F}\right)}{t-M^2}
-\sum_{I=1/2} G_{1/2}
\frac{P_J\left(1-\frac{\Sigma + s}{2\Phi}\right)}{u-M^2} \; ;
\end{displaymath}
\begin{displaymath}
-\sum_{I=1/2} G_{1/2} P_J\left(1+\frac{t}{2\Phi}\right)
\left\{\frac{1}{s-M^2}-\frac{1}{u-M^2}\right\} \; =
\end{displaymath}
\begin{equation} \label{8.1}
\; =\sum_{I=1} G_1
\frac{P_J\left(\frac{\Sigma + 2s}{4F}\right)}{t-M^2}
+\sum_{I=1/2} G_{1/2}
\frac{P_J\left(1-\frac{\Sigma + s}{2\Phi}\right)}{u-M^2} \; .
\end{equation}
>From the first of these equalities it follows that the function
$a(t)$ can be presented in the form:
\begin{equation} \label{8.2}
a(t)=
-\sum_{I=0} G_0 \frac{P_J\left(\frac{\Sigma}{4F}\right)}{t-M^2}
+\sum_{I=1/2} G_{1/2}
\frac{P_J\left(1-\frac{\Sigma}{2\Phi}\right)}{t+\Sigma} \; .
\end{equation}
Thus, the resultant expressions for the generalized amplitudes
$\overline{A}$ and $\overline{B}$ in ${\cal D}_{\overline{\nu} t}$
look as follows:
\begin{displaymath}
{\overline A}(s,t,u)=
-\sum_{I=0} G_0 \frac{P_J\left(\frac{\Sigma}{4F}\right)}{t-M^2}
+\sum_{I=1/2} G_{1/2}
\frac{P_J\left(1-\frac{\Sigma}{2\Phi}\right)}{t+\Sigma} \; -
\end{displaymath}
\begin{displaymath}
-\;\sum_{I=1/2} G_{1/2} P_J\left(1+\frac{t}{2\Phi}\right)
\left\{\left(\frac{1}{s-M^2}+\frac{1}{u-M^2}\right)+
\left(\frac{1}{M^2}+\frac{1}{\Sigma +t}\right)\right\} \; ,
\end{displaymath}
\begin{displaymath}
{\overline B}(s,t,u)=
-\sum_{I=1/2} G_{1/2} P_J\left(1+\frac{t}{2\Phi}\right)
\left\{ \frac{1}{s-M^2}-\frac{1}{u-M^2}\right\}\; ,
\end{displaymath}
\begin{equation} \label{8.3}
(s,t,u)\in {\cal D}_{{\overline \nu} t} \; .
\end{equation}

The first of the eqs. (\ref{8.3})
clearly demonstrates the absence of the fixed poles at
$t_i = - {\Sigma}_i$:
each pole originating from the correcting polynomial turns out to
be killed by the corresponding term contained in the regular
(in $s$) part $a(t)$.

With the eq. (\ref{8.2}) taken into account, the system (\ref{8.1})
reads:
\begin{displaymath}
\sum_{I=1/2} G_{1/2} P_J\left(1+\frac{t}{2\Phi}\right)
\left\{\left(\frac{1}{s-M^2}+\frac{1}{u-M^2}\right)+
\left(\frac{1}{M^2}+\frac{1}{\Sigma +t}\right)\right\} \; =
\end{displaymath}
\begin{displaymath}
=\;\sum_{I=0} G_0 \frac{P_J\left(\frac{\Sigma+2s}{4F}\right)-
P_J\left(\frac{\Sigma}{4\Phi}\right)}{t-M^2}
+\sum_{I=1/2} G_{1/2}\left\{
\frac{P_J\left(1-\frac{\Sigma +s}{2\Phi}\right)}{u-M^2}+
\frac{P_J\left(1-\frac{\Sigma }{2\Phi}\right)}{\Sigma +t}
\right\}\; ;
\end{displaymath}
\begin{displaymath}
-\sum_{I=1/2} G_{1/2} P_J\left(1+\frac{t}{2\Phi}\right)
\left\{\frac{1}{s-M^2}-\frac{1}{u-M^2}\right\} \; =
\end{displaymath}
\begin{displaymath}
\; =\sum_{I=1} G_1
\frac{P_J\left(\frac{\Sigma + 2s}{4\Phi}\right)}{t-M^2}
+\sum_{I=1/2} G_{1/2}
\frac{P_J\left(1-\frac{\Sigma + s}{2\Phi}\right)}{u-M^2} \; ;
\end{displaymath}
\begin{equation} \label{8.4}
(s,t,u)\in {\cal D}_u \; .
\end{equation}

Similar to (\ref{6.5}), one can consider (\ref{8.4})
as a system of generating equalities giving rise to an infinite set
of algebraic relations between the parameters of the resonance
spectrum. To obtain the explicit form of those relations, one has to
expand both sides of each of the eqs.  (\ref{8.4}) in a double
series in $(s-s_0)$, $(t-t_0)$ around the arbitrary point
$(s_0,t_0,u_0) \in {\cal D}_u$.

Needless to say, the comparison of (\ref{5.12}) with (\ref{6.3})
in the area ${\cal D}_s (t \leq 0 , u \leq 0)$:
\begin{displaymath}
{\cal D}_s \equiv {\cal D}_{{\overline \nu} t} \cap
{\cal D}_{{\overline t} u}\; ,
\end{displaymath}
does not provide any new information.

Now, we would like to sum up the main qualitative features of the
obtained above results. Our starting position is based on three
corner stones:
\begin{enumerate}
\item
The \underline{formal expressions} (\ref{3.10})
for the tree-level ``physical'' amplitudes $A(s,t,u)$ and $B(s,t,u)$.
\item
The \underline{asymptotic condition}
in the formulation given in Sect.4.
\item
The (intuitively justified) \underline{requirement of meromorphy}
of the ``generalized'' amplitudes $A(s,t,u)$ and $B(s,t,u)$
in a space of 3 dependent complex variables $s,t,u$.
\end{enumerate}
Further we have constructed 3 different \underline{convergent}
partial fraction expansions, each of them being suitable in its own
area ${\cal D}_{\overline{\nu}t}$, ${\cal D}_{\overline{t}s}$,
${\cal D}_{\overline{t}u}$. Since the common subdomains
${\cal D}_s$, ${\cal D}_t$, ${\cal D}_u$:
${\cal D}_s \equiv {\cal D}_{{\overline \nu} t} \cap
{\cal D}_{{\overline t} u}\; ,
{\cal D}_t \equiv {\cal D}_{{\overline t} s} \cap
{\cal D}_{{\overline t} u}\; ,
{\cal D}_u \equiv {\cal D}_{{\overline \nu} t} \cap
{\cal D}_{{\overline t} s}\;$ ,
of those areas are nonempty, we conclude that the above mentioned
expansions are mutually equal in the corresponding subdomains.

As a result, we have got 3 well defined (in terms of the spectrum
parameters) forms (\ref{6.1}), (\ref{6.3}) and (\ref{8.3})
describing the physical amplitudes $A(s,t,u)$ and $B(s,t,u)$
in the whole $(s,t,u)$-plane
(except the interior part of the Mandelstam triangle, where the
convergence is still neither postulated nor proved). The above
forms, however, are only valid subject to the conditions
(\ref{6.5}) and (\ref{8.4}),
strongly restricting the resonances parameters. From the purely
mathematical point of view, these conditions express nothing but a
requirement of analyticity. Their restrictive power arises from the
physical constraints imposed on the resulting amplitude derived
from a given form (suitable in a corresponding area) with the help
of analytical continuation.

>From the other side, the system (\ref{8.4})
clearly expresses the idea, which is commonly called as
\underline{duality}
(the equivalence of direct- and cross-channel amplitudes). The
physical origin of the conditions (\ref{6.5})
is different from that of (\ref{8.4}):
these conditions reflect the Bose symmetry requirements.

So, we conclude, that the concept of duality expresses the idea of
analytical continuation. This statement is not new. It should be
noted, however, that in the extreme form of duality (the only one
widely discussed in the literature --- see, e.g. \cite{7})
it is postulated the absence of the correcting polynomials
\underline{along with}
the absence of the nonpole (regular) term in the elastic scattering
amplitude. As it can be seen from (\ref{2.20}) and (\ref{8.3}),
this postulate gives rise to the following system of the consistency
conditions:
\begin{displaymath} \sum_{I=1/2} G_{1/2}
P_J\left(1+\frac{t}{2\Phi}\right)
\left\{\frac{1}{s-M^2}+\frac{1}{u-M^2}\right\} \; =
\end{displaymath}
\begin{displaymath}
=\;\sum_{I=0} G_0 \frac{P_J\left(\frac{\Sigma+2s}{4F}\right)}{t-M^2}
+\sum_{I=1/2} G_{1/2}
\frac{P_J\left(1-\frac{\Sigma +s}{2\Phi}\right)}{u-M^2}
\end{displaymath}
\begin{displaymath}
-\sum_{I=1/2} G_{1/2} P_J\left(1+\frac{t}{2\Phi}\right)
\left\{\frac{1}{s-M^2}-\frac{1}{u-M^2}\right\} \; =
\end{displaymath}
\begin{displaymath}
=\;\sum_{I=1} G_1 \frac{P_J\left(\frac{\Sigma+2s}{4F}\right)}{t-M^2}
+\sum_{I=1/2} G_{1/2}
\frac{P_J\left(1-\frac{\Sigma +s}{2\Phi}\right)}{u-M^2}\; ,
\end{displaymath}
\begin{equation} \label{8.5}
(s,t,u)\in {\cal D}_u\; .
\end{equation}
This system is stronger than (\ref{8.4}),
since the latter one can be derived from (\ref{8.5})
and not vice versa. Perhaps, it is too strong to describe the
reality.

Let us discuss now the another feature of the eqs.
(\ref{6.5}) and (\ref{8.4}).
As it is noted above, these conditions are equivalent to a certain
(infinite) set of algebraic relations, connecting the spectrum
parameters among themselves. In other words, they express the idea of
\underline{bootstrap}.
Of course, the complete set of bootstrap (or, the same, many-particle
duality) relations must include also those, derived from the
analyticity requirements applied to the whole set of many-particle
tree-level amplitudes.

So, we conclude, that
{\em the requirements of tree-level analyticity, duality and
bootstrap are equivalent to each other}.

This conclusion is directly related to the problem of
renormalizability of nonrenormalizable theories, discussed recently
in \cite{30}. Indeed, it is known (see, e.g., refs.
\cite{31,32,33,34,35,36,37,38,39}),
that on the tree-level the renormalizability requirement corresponds
to that of the polynomial boundedness of an amplitude, which is
implied to be a meromorphic function. Since the tree-level amplitude
determines the Lagrangian of a theory, the above conclusion can be
formulated as follows: one has no chance to construct a
renormalizable theory (with unbounded spectrum of mass
and spin)  until the bare triple (on-shell) coupling constants and
bare masses are restricted by the infinite set of the analyticity
requirements. (As it follows from our analysis, all $n$-particle
couplings with $n \geq 4$ can be expressed
in terms of the above mentioned parameters.)

\section{Chiral expansion coefficients and resonance saturation.}
\mbox{}

In this section we would like to analyze the role of chiral
symmetry in the discussed above dual picture of tree level hadron
interactions.

Let us consider the coefficients $b_{ij}$ of the amplitude $B(\nu,t)$
power series expansion around the point $\nu =0,\; t=0$ :

\begin{equation} \label{9.1}
B(\nu,t)=\nu\sum_{i,j=0}^{\infty}b_{ij}\nu^{2i}t^j \; .
\end{equation}
In the case of ${\mu}^2 = 0$ (massless pion), the eq.
(\ref{9.1}) is nothing but the Chiral expansion
\cite{40} written in the large-$N_c$ limit,
$b_{ij}$ being the linear combination of corresponding ChPT expansion
coefficients defined in \cite{41} (for a review see, e.g.,
\cite{42}).
Let us consider first the lowest coefficient
$b_{00}$. With the help of (\ref{5.12}) one obtains

\begin{equation} \label{9.2}
b_{00}=\sum_{I=1/2}G_{1/2}\frac{1}{(M^2 -m^2 )^2} \; .
\end{equation}
>From the other side, the Chiral symmetry tells us that
\begin{equation} \label{9.3}
b_{00}=\frac{1}{4{f_0}^2} \; ,
\end{equation}
where $f_0\approx 87 Mev.$
Thus we conclude that the following sum rule (SR) holds:
\begin{equation} \label{9.4}
\frac{1}{4{f_0}^2}=\sum_{I=1/2}G_{1/2}\frac{1}{(M^2 -m^2 )^2}
\end{equation}

The RHS of
(\ref{9.4}) should be computed with the values of masses and coupling
constants taken at
${\mu}^2 =0$  (chiral limit). As it has been shown in ref.
\cite{2}, the corresponding SR for
$\pi\pi$--scattering is correct to the accuracy of experimental data.
However, (see \cite{9}), the detailed numerical analysis of
(\ref{9.4}) would be premature, since the current information on
$\pi K$--system is still too scarce.

Here we would like to emphasize another --- purely theoretical ---
role of sum rules of the type
(\ref{9.4}). Considering the consistency conditions (\ref{6.5}) and
(\ref{8.4}), one notes that they are homogeneous with respect to the
coupling constants
$G_I$. Indeed, one can simultaneously multiply all
$G_I$  in the above conditions by the same arbitrary factor
$S$ without breaking the equality. This is not true for the eq.
(\ref{9.4}). So, we conclude that
{\em the Chiral Symmetry provides a unique normalization scale for
coupling constants}
$G_I$. It is not difficult to understand that this conclusion is
valid also for every many-particle process incorporating two pions.

The form (\ref{5.12}) of the elastic scattering amplitudes
$A(s,t,u)$ and
$B(s,t,u)$ provides also the natural justification for the effect of
so-called ``resonance saturation'', namely, the agreement between the
phenomenologically determined values of 4-th order ChPT expansion
coefficients and the magnitudes given by the contributions of the
relevant low-lying resonances (see
\cite{41,43,44}). Indeed, if one computes the value of a given
coefficient with the help of
(\ref{5.12}), he obtains sum rule similar to
(\ref{9.2}) --- the corresponding examples are considered in detail
in refs.
\cite{1,2,8,9,10} . Each of those SR takes the form of a sum over the
resonance contributions, the most significant terms corresponding, as
a rule, to the lowest resonances. Thus one obtains the natural
explanation for the effect of resonance saturation.

There is, however, the essential difference between our approach
and that considered in ref.
\cite{41,43,44}, this difference appearing already in the chiral
limit ${\mu}^2 =0$. The thing is that we do not need to use
any special --- chiral --- form for triple vertices: the difference
between chiral and non-chiral couplings is assigned to the regular
(non-pole) part of the
amplitude. The latter one, in turn, is completely determined
by the analyticity requirement. In our approach the chiral symmetry
manifests itself in the values of masses and on-shell coupling
constants. In other words, in order to obey the tree level
analyticity requirements, the chiral symmetry must be realized as
an ordinary \underline{algebraic} symmetry (we use the terminology
suggested in ref.
\cite{3}). The similar conclusion (in a much more strong form)
has been first drawn by Weinberg
\cite{3,15} from the analysis of the asymptotic condition
at zero momentum transfer.

It should be noted also, that the importance of the asymptotic
requirement has been pointed out in ref.
\cite{45a} , where the authors discuss various schemes of
accounting for the
vector meson contribution. Therefore, it is interesting to
compare in more detail two different approaches to the problem of
resonances in the framework of ChPT: our one and that suggested in
\cite{43,44}.

Let us consider, for simplicity, the ``$SU_2$ chiral world''
(${\mu}^2 =0,\: m^2 \ne 0$) in the large-$N_c$
limit. At very small values of the pion CMS momentum
$q$ , the ChPT expansion provides the
{\em most general form} of the
$\pi K$ -amplitude consistent with the QCD requirements. Therefore,
{\em this form contains already all the information on the
corresponding resonances.} When constructed, it completely
determines the amplitude to a given accuracy, which can be taken
arbitrarily high. What happens with increasing
$q$ ? In the first stage
($s<M_{K^\ast}^2$) nothing terrible happens, since, to provide the
given level of accuracy, one can add new and new terms of higher
orders. However, at
$s=M_{K^{\ast}}^2$ the expansion diverges, because in this
point the amplitude has a pole. Hence, one has to reorganize the
chiral expansion
{\em in a self-consistent manner}, allowing to isolate the
$K^{\ast}$ -pole
\underline{explicitly}. Here the term ``self-consistent'' means that
at $s<M_{K^{\ast}}^2$ two expansions --- old and new --- must
coincide identically.

The corresponding method is known from the potential scattering
theory
\cite{45,47,48}. One has to introduce the new particle into a theory
and, simultaneously, change the potential. When applied to the case
in question, this prescription can be best illustrated by the
following equality:
\begin{equation} \label{9.5}
\sum_{k=0}^{\infty}a_k q^{2k}
=\sum_{k=0}^{\infty}b_k q^{2k}
+\frac{r}{q^2 -P} \;\;\; ,
\end{equation}
where $q^2<P$ and

\begin{equation} \label{9.6}
b_k = a_k + \frac{r}{P^{2k+1}} \; ,
\end{equation}
both $r$ and $P$ being constants.
\underline{By suggestion}, the LHS of (\ref{9.5}) converges at
$q^2<P$ . In contrast, the convergence area of a sum in the RHS is
some wider: it is bounded by the next resonance position
$P'>P$ . So, the eq.
(\ref{9.5}) can be used for the analytical continuation of the
\underline{chiral} expansion appearing in its LHS, the chiral
symmetry of the resulting expression being guaranteed
\underline{by construction},
{\em regardless of the particular form of $\pi KR$ -vertex.}

Repeating the above procedure step by step, one obtains the
$\pi K$ -amplitude in the form used in previous Sections, with all
resonances being explicitly taken into account.

It should be noted, that the analytical continuation of chiral
expansion could be based (with equal success) on the equality

\begin{equation} \label{9.7}
\sum_{k=0}^{\infty}a_k q^{2k} =\sum_{k=0}^{\infty}c_k q^{2k}
+\frac{\sum_{k=0}^{N}r_k q^{2k}}{q^2 -P} \;
\end{equation}
with arbitrary \underline{finite}
$N$ . This means, that one can use the
$\pi KR$ -vertices with arbitrary finite number of derivatives.
However, one should exercise an extreme caution when taking limit
$N\to\infty$ , since the meromorphy of the resulting form of the
amplitude cannot be guaranteed in this case without special efforts.
This note shows that it makes not so much sense to organize the power
counting for $\pi KR$ -vertices; such a counting is only sensible for
$q^2\ll M_{K^{\ast}}^2$ .

Now, the difference between two approaches under consideration can be
easily understood. Indeed, the only distinctive feature of chiral
couplings --- comparing to the minimal ones --- consists in the
number of derivatives acting on the pion field. As it is explained in
Sec.2, this difference results in the appearance of extra terms in
the numerators of the resonance
propogators. This, in turn, means that at every
\underline{fixed} order of chiral expansion the approach of
\cite{43,44} is completely equivalent to that considered
in this paper; it is not difficult to establish the one-to-one
correspondence between the coefficients
$c_k , r_k$ appearing in (\ref{9.7}) and $b_k$ given by
(\ref{9.6}). At the same time, our approach looks preferable, since
it guarantees one that no unwanted singularities can appear in the
process of analytical continuation of chiral expansion.

Here it is pertinent to note that the commonly met statement on the
arbitrariness of chiral expansion coefficients is nothing but a
misunderstanding. It is true that Chiral Symmetry tells us
nothing about their values. However, one should not forget that the
structure of
\underline{nonlocality} of Effective Chiral Lagrangian (infinite
number of derivatives!) is by no means arbitrary. This structure
stems from a certain procedure of ``integration out'' of all
``heavy'' degrees of freedom. In the large-$N_c$
limit it becomes especially transparent, since in this case ---
as we believe --- the only possible degrees of freedom are colorless
hadrons
\cite{4,5} . Therefore, it is not surprising that the values of the
chiral expansion coefficients are connected with the resonance
spectrum parameters.

\section{Concluding remarks.}
\mbox{}

In this section we give a brief summary of the most interesting of
our results and point out some open questions.

Perhaps, the most interesting result consists of demonstration
of a power of Weinberg's asymptotic condition, formulated as a
tree-level analyticity requirement (meromorphy and polynomial
boundedness). This very requirement, which looks trivial in the
case of a system with the finite number of resonances, made it
possible for us: 1) To \underline{prove}
the duality in its most general form; 2) To formulate the system
of bootstrap equations; 3) To demonstrate the equivalence of the
bootstrap and duality conditions.

This result, in fact, provides a general solution of the problem
of dispersion relations saturation with one-particle states. Of
course, it would be very interesting to put it into the
algebraic form (similar to that obtained by Weinberg in refs.
\cite{3,15}). This work is in progress. It should be stressed,
however, that the given above form (\ref{8.3})
is quite sufficient for the purely practical needs. In particular,
it gives one a possibility to estimate the chiral expansion
coefficients with the accuracy provided by the experimental data.

One more result, which we would like to mention, concerns with a
possibility to construct a renormalizable theory of higher spin
particles (the ``renormalization problem for nonrenormalizable
theories'' --- see \cite{30}).
Indeed, it can be shown, that any given graph, written in the
conventional terms (propagators in the ``unitary'' form and
arbitrary vertices), can be rewritten in our ones (on-shell
vertices and on-shell propagators, given by the first term of
(\ref{2.1}) plus series of graphs with less number of loops, on-shell
propagators and renormalized triple (on-shell) couplings, the latters
obeying the duality requirements. So, the problem of
renormalizability (or, better to say, finiteness) looks quite similar
to that, already considered in the framework of known dual schemes
(see, e.g., \cite{49}).

There are some open questions which it would be interesting to study
in more detail:
\begin{enumerate}
\item
It looks interesting to formulate the necessary and sufficient
conditions, providing the convergence of the considered above
partial fraction expansions inside the Mandelstam triangle. Perhaps,
this could give us the more detailed information on the structure of
spectrum (leading trajectories, satellites, something else?). The
results of refs. \cite{50,51} support this idea.
\item
It would be interesting to answer the question on the form
(\ref{8.3})
consistency with the tree-level unitarity condition:
$|\mbox{Re}\: a_l|\leq 1/2\; .$
(here $a_l$ denote the suitably normalized $l$-th
partial amplitude). The importance of this condition for the
understanding of the low-energy $\pi \pi$
scattering has been demonstrated recently in \cite{52,53}.
\item
Possibly, it would be interesting to look for the solution of the
bootstrap conditions, using the method of ref. \cite{54}.
\item
As we have already pointed out above, the closed algebraic form of
the bootstrap conditions would be of great interest as well as the
form of the corresponding Lagrangian.
\end{enumerate}

To conclude, we would like to stress, that the asymptotic conditions
--- regardless to their particular form --- give one the powerful
tool for the understanding of the hadron spectrum structure. The
results of the recent papers \cite{55,56}
give the further argument in favor of this statement.

\section{Acknowledgments.}
\mbox{}

It is a pleasure to thank my colleague M.Polyakov for his friendly
support and conversations. I am indebted to A.Vereshagin for
numerous stimulating discussions of the complex analysis problems
and for the help in preparation of the manuscript. I thank
also A.Bramon and M.Eides for discussions.

This work was supported in part by RFFI (Grant 96-02-18017) and by
GRACENAS (Grant 95-06.3-13).


\begin{thebibliography}{57}
\bibitem{1} A. Bolokhov, A. Manashov, M. Polyakov, V. Vereshagin.
 Phys. Letters,
\underline{B303} , 220 (1993).
\bibitem{2} A. Bolokhov, A. Manashov, M. Polyakov, V. Vereshagin.
 Phys. Rev.,
\underline{D48} , 3090 (1993).
\bibitem{3} S. Weinberg. Phys. Rev.,
\underline{177} , 2604 (1969).
\bibitem{4} G. t'Hooft. Nucl. Phys.,
\underline{B72} , 461 (1974).
\bibitem{5} E. Witten. Nucl. Phys.,
\underline{B160} , 57 (1979).
\bibitem{6} G. Veneziano. Phys. Reports,
\underline{9C} , 199 (1974).
\bibitem{7} M. Fukugita, K. Igi. Phys. Reports,
\underline{31} , 237 (1977).
\bibitem{8} A. Bolokhov, M. Polyakov, V. Vereshagin.
 Sov. J. Nucl. Phys.,
\underline{53} , 251 (1991).
\bibitem{9} A. Bolokhov, A. Manashov, M. Polyakov, V. Vereshagin.
 hep-ph/9503424 (1995).
\bibitem{10} M. Polyakov, V. Vereshagin.
 Preprint PNPI TH-41-1995 2068.
 hep-ph/9509259 (1995). To be published in Phys. Rev. D.
\bibitem{11} V. Vereshagin. Nucl. Phys.,
\underline{B55} , 621 (1973).
\bibitem{12} S. Weinberg. Phys. Rev.,
\underline{133B} , 1318 (1964);
\underline{134B} , 882 (1964);
\underline{181} , 1893 (1969).
\bibitem{13} M. D. Scadron. Phys. Rev.,
\underline{165} , 1640 (1968).
\bibitem{14} A. Bolokhov, A. Manashov, M. Polyakov, V. Vereshagin.
 Phys. Rev.,
\underline{D50} , 4713 (1994).
\bibitem{15} S. Weinberg. Phys. Rev. Letters,
\underline{65} , 1177 (1990).
\bibitem{15a} H. Harari. Phys. Rev. Letters,
\underline{20} , 1395 (1968).
\bibitem{16} V. de Alfaro, S. Fubini, G. Furlan, C.Rossetti.
 Currents in Hadron Physics. North--Holland Publishing Co.,
 Amsterdam--London, 1973.
\bibitem{17} C. Fronsdal. Nuovo Cimento Suppl.,
\underline{9} , 416 (1958).
\bibitem{18} P. L. Csonka, M. J. Moravcsik, M. D. Scadron.
 Annals of Physics,
\underline{40} , 100 (1966).
\bibitem{19} A. Bolokhov, N. Kivel, M. Polyakov, V. Vereshagin.
 Preprint FUB-HEP/90-22, Berlin, 1990.
\bibitem{20} A. Bolokhov, V. Vereshagin. Sov. J. Nucl. Phys.,
\underline{26} , 588 (1977);
\underline{28} , 133 (1978);
\mbox{\underline{29} ,} 97 (1979);
\underline{29} , 670 (1979).
\bibitem{21} V. I. Ogievetsky, B. M. Zupnik. Nucl. Phys.,
\underline{B24} , 612 (1970).
\bibitem{22} V. I. Ogievetsky. Phys. Letters,
\underline{33B} , 227 (1970).
\bibitem{23} V. I. Ogievetsky. Sov. J. Nucl. Phys.,
\underline{13} , 105 (1971).
\bibitem{24} B. Zumino. Hadrons and Their Interactions.
             NY, 1968. p. 51.
\bibitem{25} B. Zumino. Theory and Phenomenology in Particle Physics.
 NY, 1969. p. 42.
\bibitem{26} C. Cr\"{o}nstorm, M. Noga. Nucl. Phys.,
\underline{B15} , 61 (1970).
\bibitem{27} C. Cr\"{o}nstorm, M. Noga. Phys. Rev.,
\underline{D1}, 2414 (1970).
\bibitem{28} V. I. Ogievetsky. In '' Essays Dedicated on the Occasion
 of the 75-th Birthday of \mbox{I. E. Tamm}''. Moscow, 1972.
\bibitem{29} A. Sudbery. Nucl. Phys.,
\underline{B20} , 1 (1970).
\bibitem{30} J. Gomis, S. Weinberg. Preprint UTTG-18-95, 1996.
\bibitem{31} S. Weinberg. Phys. Rev. Letters,
\underline{27} , 1688 (1971).
\bibitem{32} A. Vainstein, I. B. Khriplovich. Sov. J. Nucl. Phys.,
\underline{13} , 111 (1971).
\bibitem{33}  J. M. Cornwall, D. N. Levin, G. Tiktopoulos.
 Phys. Rev. Letters,
\underline{30} , 1268 (1973).
\bibitem{34} J. M. Cornwall, D. N. Levin, G. Tiktopoulos.
 Phys. Rev.,
\underline{D10} , 1145 (1974).
\bibitem{35} D. N. Levin, G. Tiktopoulos. Phys. Rev.,
\underline{D12} , 415 (1975).
\bibitem{36} C. H. Llewellyn Smith. Phys. Letters,
\underline{46B} , 233 (1973).
\bibitem{37} J. S. Bell. Nucl. Phys.,
\underline{B60} , 427 (1973).
\bibitem{38} J. Schechter, Y. Ueda. Phys. Rev.,
\underline{D7} , 3119 (1973).
\bibitem{39} S. Joglekar. Annals of Physics,
\underline{83} , 427 (1974).
\bibitem{40} S. Weinberg. Physica,
\underline{96A} , 327 (1979).
\bibitem{41} J. Gasser, H. Leutwyler. Annals of Physics, NY,
\underline{158} , 142 (1984).
\bibitem{42} A. Pich. Reports on Progr. in Physics.
\underline{58} , 563 (1995).
\bibitem{43} G. Ecker, J. Gasser, A. Pich, E. de Rafael. Nucl. Phys.,
\underline{B321} , 311 (1989).
\bibitem{44} J. F. Donoghue, C. Ramirez, G. Valencia. Phys. Rev.,
\underline{D39} , 1947 (1989).
\bibitem{45a} G. Ecker, J. Gasser, H. Leutwyler, A. Pich,
E. de Rafael.
Phys. Letters,
\underline{223B} , 425 (1989).
\bibitem{45} S. Weinberg. Phys. Rev.,
\underline{130} , 776 (1963);
\underline{131} , 440 (1963).
\bibitem{47} M. Scadron, S. Weinberg. Phys. Rev.,
\underline{133} , B1589 (1964).
\bibitem{48} M. Scadron, S. Weinberg, J. Wright. Phys. Rev.,
\underline{135} , B202 (1964).
\bibitem{49} M. B. Green, J. H. Schwarz, E. Witten.
 Superstring Theory. Cambridge Univ. Press, Cambridge, 1987.
\bibitem{50} C. Lovelace. Phys. Letters,
\underline{28B} , 264 (1968).
\bibitem{51} M. Ademollo, G. Veneziano, S. Weinberg.
             Phys. Rev. Letters,
\underline{22} , 83 (1969).
\bibitem{52} F. Sannino, J. Schechter. Phys. Rev.,
\underline{D52} , 96 (1995).
\bibitem{53} M. Harado, F. Sannino, J. Schechter.
Preprint SU-4240-624, 1995, hep-ph/9511335.
\bibitem{54} A. Bramon, E. Masso. Lettere al Nuovo Cimento,
\underline{20} , 621 (1977).
\bibitem{55} A. Wirzba, M. Kirchbach, D. O. Riska.
Journ. of Phys. G,
\underline{20} , 1583 (1994).
\bibitem{56} R. Dashen, E. Jenkins, A. Manohar. Phys. Rev.,
\underline{D49} , 4713 (1994).
\end{thebibliography}
\end{document}